\documentclass[12pt]{article}
\usepackage{graphicx}
\begin{document}
{\it For APFA3, London, Dec. 2001}
\bigskip

{\bf Nucleation of Market Shocks in Sornette-Ide model}

\bigskip

Ana Proykova$^1$, Lena Roussenova$^2$ and Dietrich Stauffer$^3$

$^1$ Department of Atomic Physics, University of Sofia, Sofia-1126, Bulgaria

$^2$ Department of Finance, University of National and World Economy, Sofia,
Bulgaria

$^3$ Institute for Theoretical Physics, Cologne University, D-50923 K\"oln, Euroland

\bigskip

Abstract:
The Sornette-Ide differential equation of herding and rational trader behaviour
together with very small random noise is shown to lead to crashes or bubbles
where the price change goes to infinity after an unpredictable time. About 100 
time steps before this singularity, a few predictable roughly log-periodic 
oscillations are seen.

\bigskip
Bubbles and crashes are a property of market prices since centuries. Some may
arise from external perturbations like changes of the interest rate by the 
central bank, government decisions on war and peace, or for stocks of single
companies the introduction of new (un)successful products. Some, like the
October 1987 crash on Wall Street, may arise from intrinsic market mechanisms
[1]. While we cannot make a clear distinction between external and intrinsic 
reasons for a crash in reality, we can do so at least in computer simulations.
There a single external event can be put in explicitly into the program,
the multitude of individual decisions can be approximated by a small 
random noise while the general rational as well as psychological (``herding'')
behaviour of investors and traders can be approximated by deterministic 
algorithms. We show here that in the Sornette-Ide [2,3] approach this small 
noise can escalate to a singularity (bubble or crash) where the price change
goes to $\pm \infty$.

The Sornette-Ide approach [2] used two different contributions. One is
the nonlinear fundamentalist assumption of
[3] that traders buy (sell) if the prices are low (high). Thus a positive
difference $x$ between the actual and the perceived fundamental price encourages
selling and thus causes a downwards tendency in $v = dx/dt$ such that $dv/dt$ 
is proportional to $-x^n$. This effect stabilizes prices. But a rising price,
$v > 0$, creates the hope or illusion that the price will increase further 
and thus will encourage further buying, with $dv/dt$ varying as $v^m$; this
second term alone leads to a divergence $v \propto (t_c-t)^{-1/(m-1)}$ at some
critical time $t=t_c$ (bubble or crash). The combination of both terms leads to
roughly [4] log-periodic oscillations $\propto \sin[{\rm const}\,\log(t_c-t)]$
preceding the singularity, which may be used to make profit or to prevent this 
singularity. 

\begin{figure}[hbt]
\begin{center}
\includegraphics[angle=-90,scale=0.5]{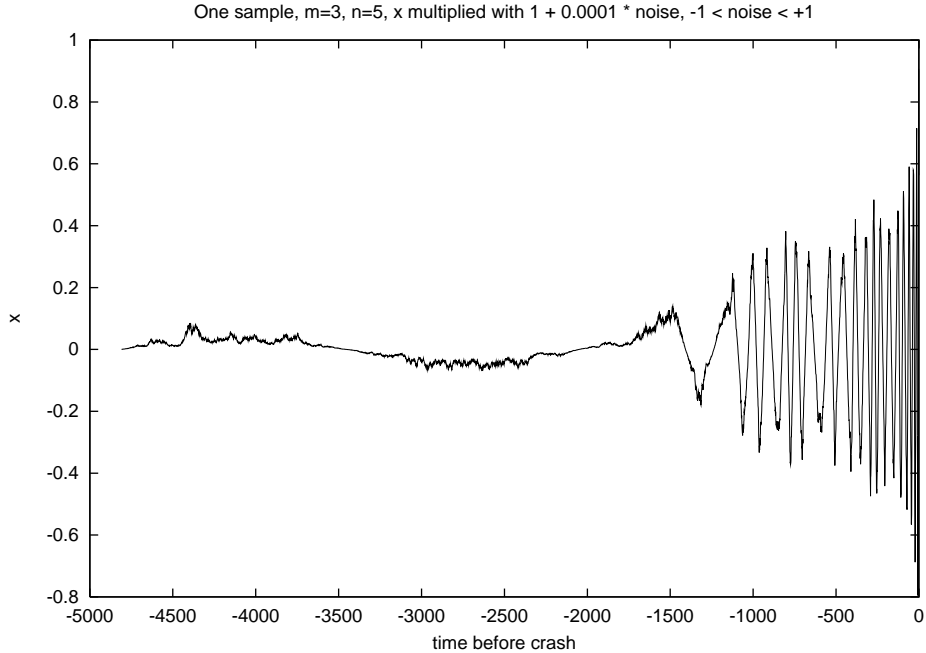}
\end{center}
\caption{
Long-time behaviour of one example. Slow weak oscillations are followed
by strong rapid oscillations.
}
\end{figure}

\begin{figure}[hbt]
\begin{center}
\includegraphics[angle=-90,scale=0.5]{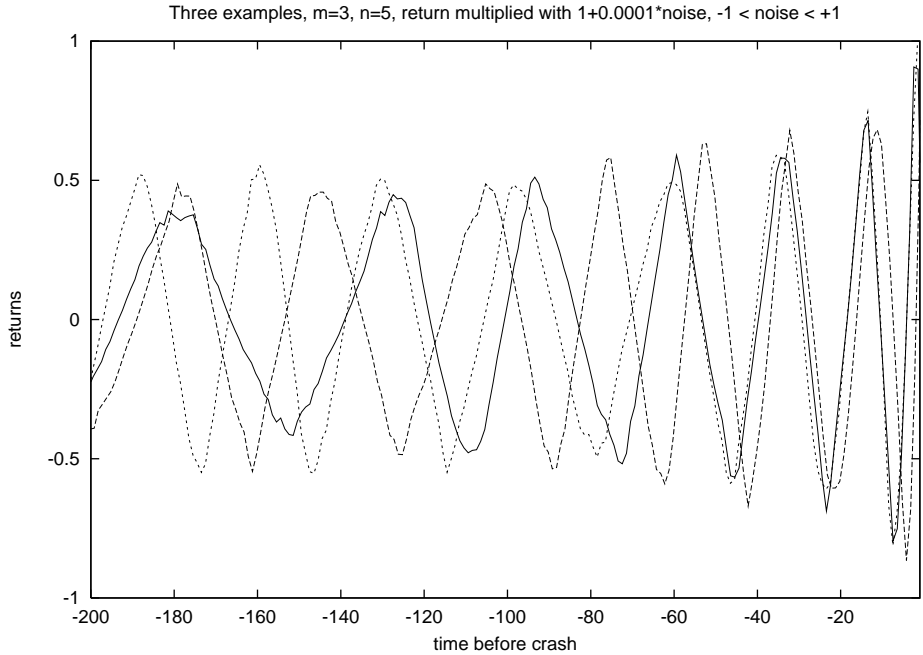}
\end{center}
\caption{
Behaviour shortly before the crash of the same sample as in Fig.1 plus
two other examples differing only by the random number seed. Only for a few
oscillations the three curves are reasonable synchronized; for longer times
before the crash some show a maximum where the others show a minimum.
}
\end{figure}

\begin{figure}[hbt]
\begin{center}
\includegraphics[angle=-90,scale=0.5]{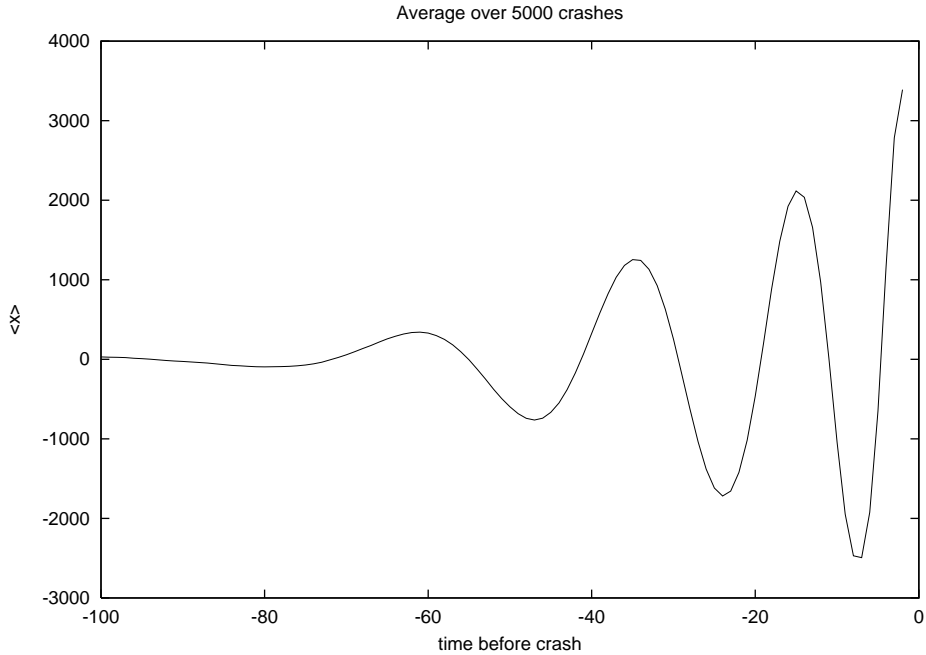}
\end{center}
\caption{
Average over many crashes, for the same parameters as in Figs.1,2.
}
\end{figure}

\begin{figure}[hbt]
\begin{center}
\includegraphics[angle=-90,scale=0.5]{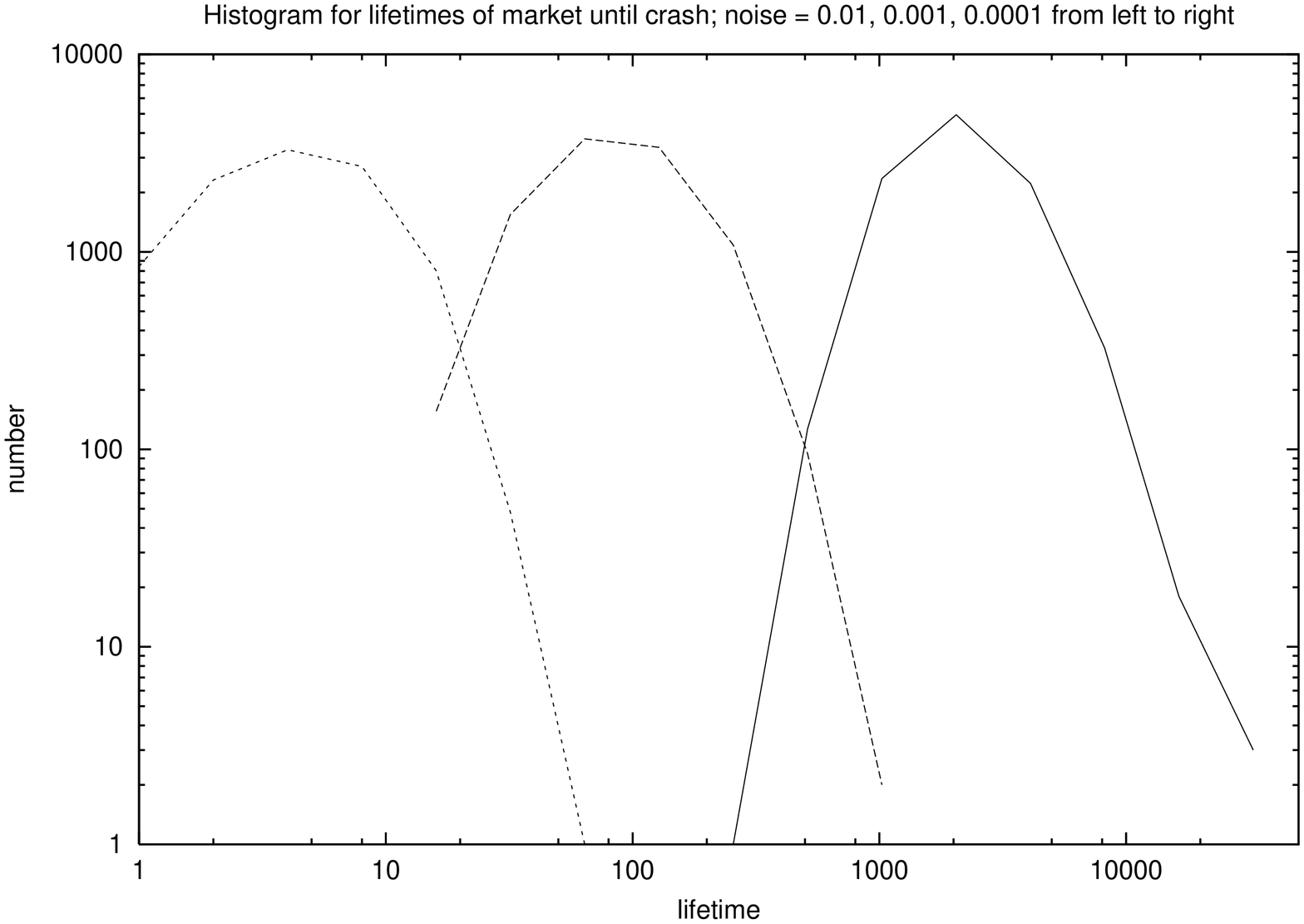}
\end{center}
\caption{
Distribution of the times after initialization needed for a crash to
appear. A parabola in this log-log plot means a log-normal distribution. The
bin sizes increase by a factor 2. Besides our normal noise 0.0001 on the right, 
also ten and hundred times stronger noise is shown (center and left curves.) 
}
\end{figure}

The deterministic nonlinear differential equation for the ``return'' $x(t)$ (the
logarithm of the current price $P$ to the initial ``fundamental'' or 
``equilibrium'' price) and its velocity (short-time return) $v = dx/dt$ is
$$ dv/dt = c_mv^m - c_nx^n $$
if $m$ and $n$ are odd integers. The damped harmonic oscillator has $m=n=1,\; 
c_m < 0, \;c_n > 0$. Suitable absolute values are needed if $m$ and $n$ are
not odd integers [2]. Noise can be introduced here in various ways: Chang et al 
[5] added a small random value to $x$; then after a long time regular 
oscillations appeared, of increasing strength and decreasing period, until $x$ 
diverged. While this behaviour is what we want, the assumption about the noise 
has little justification except that it is close to Langevin equations.  

We instead first made $c_m$ 
vary randomly between 0 and +1, but then the curves were too regular,
and different initial seeds for the random number generator gave nearly the
same curves. This multiplicative assumption was supposed to simulate the
volatile psychology of the traders who follow more or less the current trend.
When we applied the random noise to $c_n$ instead of $c_m$ this noise influenced
the prices more strongly but we got oscillations right from the beginning 
while we want them to emerge only later. Also applying noise to $v$ did
not give satisfactory results. 

To get better results, in our second method 
we made a multiplication for $x$ itself, not for
a contribution to its time derivatives. Thus the $x$ resulting from the
discretized differential equation was multiplied by a factor which differed
from unity by a small amount between --0.0001 and +0.0001. Since we use
multiplication instead of addition, we cannot start with $x$ and $v$ both
zero, and thus took $x=0, \; v = 0.0001$ initially, to create some fluctuations.
We thus assume that traders decide in a deterministic way on fundamental value 
through $c_nx^n$ and on herding through $c_mv^m$, but finally some small
randomness also affects the market.

Figure 1 shows a run over all nearly 5000 time steps, Fig.2 is the same run over
the last 200 time steps before the crash, together with two other runs using 
the same parameters but different random numbers. We see that first the prices 
fluctuate in a rather random way about the fundamental value.  In other runs, 
the noise can make the deviations larger leading to a few nearly 
regular oscillations which die down again. Finally, again such oscillations 
occur, but now they become stronger, faster and finally lead to a crash at time = 0. Averaging over thousands of samples gives smooth oscillations before
the crash, Fig.3. If we increase the noise, the times needed for a crash to
build up get shorter and there are less oscillations; these times are roughly
log-normally distributed, Fig.4. 

Mathematically it is easy to understand why this second method works better
than our first method. In the first method, the noise affects the time
derivative only, meaning it changes the curvature or slope of $x(t)$ only. Our 
second method affects $x$ directly.

Without any noise, the origin $x=v=0$ of the $(v,x)$ phase space plays a special
role as the unstable fixed point around which spiral structures of trajectories
are organized. It corresponds to the case of fundamental prices; there is no
trend and the market does not know which direction to take. 

Our parameters for this simulation were $m=3,\; n=5,\; c_m=c_n=1$. We used the
leap frog method of molecular dynamics with 1000 iterations per unit time.
Thus first we calculate from the above equation the acceleration $a = 
d^2x/dt^2$, from $a$ the change $0.001 a$ in the velocity $v = dx/dt$, 
and finally the new $x$ is obtained by
first adding $0.001 v$ to it using the new velocity, and then multiplying 
$x$ by $1 + \epsilon$, with $\epsilon$ selected randomly between --0.0001 and
+ 0.0001. 

The ``noise'' $\epsilon$ symbolizes all the new information coming in every day
(or every simulation interval). We also tried to simulate insider trading: Half 
of the market movement uses not today's noise but the one of tomorrow. Not much
is changed since the equations describe the overall market behaviour, not the
profits of some at the expense of others.

Of course, this method is similar to that of Chang et al; the main difference 
is that the present noise in $x$ is proportional to $x$ while for Chang et al
it is independent of $x$. 
An additional linear friction term, $ dv/dt = c_mv^m - c_nx^n -c_f v, $
restricts the fluctuations $x$ mostly to one sign, except for very small
friction coefficients $c_f$.

\bigskip
We thank D. Sornette for helpful criticism.
This work was supported by the Sofia-Cologne university partnership.
\parindent 0pt

\bigskip
\bigskip
[1] G.W. Kim and H.M. Markowitz, J. Portfolio Management 16, 45 (fall 1989).

[2] D. Sornette and K. Ide, cond-mat/0106054; K. Ide and D. Sornette, 
cond-mat/0106047

[3] R.B. Pandey and D. Stauffer, Int. J. Pure Appl. Finance 3, 279 (2000)

[4] D. Sornette and A. Johansen, Quant. Finance 1, 457 (2001).

[5] I. Chang, D. Stauffer and R. B. Pandey, cond-mat/0108345 (2001).

\end{document}